\documentclass[a4paper,longbibliography,twocolumn,superscriptaddress,pre, aps, floatfix,longbibliography ]{revtex4-1}

\usepackage[utf8]{inputenc}
\usepackage[english]{babel}
\usepackage{amssymb}
\usepackage{amsmath}
\usepackage{graphicx}
\usepackage{float}
\usepackage[normalem]{ulem}
\usepackage{color}

\newcommand{\eqcomma}{\,,}

\usepackage[percent]{overpic}

\newcommand{\piclabt}[2]{
\begin{overpic}[width=0.24\textwidth]{#1}
 \put (3,95) {\large \textsf{#2)}}
\end{overpic}
}

\begin{document}

\author{Roberta Amato}
\author{Albert D\'iaz-Guilera}
\affiliation{Departament de Fisica de la Matèria Condensada, Marti i Franques 1, Universitat de Barcelona, 08028 Barcelona, Spain}
\affiliation{Institute of Complex Systems (UBICS), Universitat de Barcelona, Mart\'i i Franqu$\grave{e}$s 1, E-08028, Barcelona, Spain}
\author{Kaj-Kolja Kleineberg}
\email{kkleineberg@ethz.ch}
\affiliation{
Computational Social Science, ETH Zurich, Clausiusstrasse 50, CH-8092 Zurich, Switzerland}

\date{\today}

\title{Interplay between social influence and competitive strategical games in multiplex networks}

\begin{abstract}
We present a model that takes into account the coupling between evolutionary game dynamics and social influence.
Importantly, social influence and game dynamics take place in different domains, which we model as different layers of a multiplex network. 
We show that the coupling between these dynamical processes can lead to cooperation in scenarios where the pure game dynamics predicts defection. 
In addition, we show that the structure of the network layers and the relation between them can further increase cooperation. 
Remarkably, if the layers are related in a certain way, the system can reach a polarized metastable state.
These findings could explain the prevalence of polarization observed in many social dilemmas.
\end{abstract}

\maketitle

\section*{Introduction}

Social dilemmas are situations where individual interests are in conflict, like sharing resources or generating common goods. 
These situations are commonly modeled by the Prisoner's Dilemma~\cite{rap1965} or the Stag Hunt game. 
Remarkably, cooperation in 
situations where individual interest are in conflict
is surprisingly common in reality, although defection has been shown to prevail in many theories and controlled experiments~\cite{gru2012,gru2014,fle2009}.
Several mechanisms have therefore been proposed to explain the emergence of cooperation in these scenarios, 
for example direct and indirect reciprocity (image scoring/reputation)~\cite{jac11}, kin and group selection~\cite{Nowak2006,Nowak1998}, success-driven migration~\cite{Helbing2009pnas}, or punishment \cite{boyd2003}.

Another mechanism responsible for the emergence of cooperation in social dilemmas could be the fact that strategic interactions between individuals or institutions do not occur in isolation. In particular, individuals that engage in strategic interactions are simultaneously exposed to social influence and, consequently, the spread of opinions. 
Following this line of reasoning, we assume that social influence 
impacts the decisions of the players~\cite{Vilone2012,PhysRevE.90.022810}, and that, vice versa, the decisions of the players impact the opinions that are propagated in the system. 
This consideration naturally raises the following questions:
Can the interplay between social influence and game theoretical decisions enable cooperation in social dilemmas? And, what is the impact of the network topology and the relation between the structure of the social and strategical domain?

To answer these questions, we present a model where game theoretical decisions and social influence take place in different layers of a multiplex network~\cite{kiv14, per2016,gom15,gom2012, mat2015, san2014, Wang2015}. In such systems, each layer contains the same set of nodes, but links are usually different in different layers. However, the layers comprising real multiplex systems are not entirely independent, but exhibit certain relations~\cite{geometry:multilayer}. As we will show, these relations can lead to interesting behaviors, and hence have to be taken into account when modeling such systems~\cite{geo:targeted}. 
On top of this topology, we model the dynamics taking place in the game layer by a replication dynamics, where individuals imitate the strategy of successful neighbors~\cite{evogamesgraphs,Cressman22072014,Helbing1996rep}. Furthermore, we use a biased voter model~\cite{opac-b1095541} in the social influence layer as a proxy for the spread of opinions. These opinions can be seen as a proclamation of the intend of individuals regarding their choice of strategy in the game layer. We assume that there is a tendency for individuals to act in agreement with their proclamations, but allow, in general, that individuals deviate from them. The importance for individuals to be congruent in both domains constitutes the coupling strength between the different dynamical processes. Finally, the aforementioned bias of the voter model represents a general tendency towards the proclamation of cooperative intentions, which could be induced by appropriate media campaigns or similar measures.

We find that indeed the coupling between evolutionary game dynamics and social influence can sustain partial cooperation in the Prisoner's Dilemma, 
where total defection prevails in the isolated game dynamics, 
and partial or even full cooperation in the Stag Hunt game.
In both cases, for appropriate parameters, the state of full defection can be avoided entirely. 
The role of the relation between the different layers of the system is especially interesting. In particular, 
only certain relations between the layers give rise to a metastable state in which nodes that adopt the same strategy self-organize into local clusters. This state could explain the emergence and prevalence of polarization observed in many situations that resemble social dilemmas. 

\begin{figure}[t]
\centering
\includegraphics[width=1\linewidth]{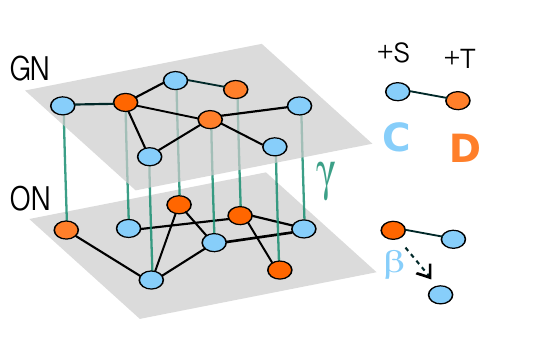} 
 \caption{Graphical illustration of the multiplex model. Different layers denote the different networks that
individuals participate: on the top the Game Network (GN) and on the bottom the Opinion Network (ON). Intra-links (black solid lines) correspond to the individuals' connections within each network, while
inter-links (green solid lines) indicate the coupling between layers. On the right, the pictures show a simplification of the dynamic occurring on each network.
\label{mod}
}
\end{figure}
\section*{Results}
\subsection*{Coupling between game dynamics and social influence}

In strategical games, 
individuals choose a strategy and then obtain a payoff that depends on their own and other players' strategies.
Here, we consider the two possible strategies cooperate (C) or defect (D). 
The interactions are then governed by the payoff matrix
\begin{equation}
M=
\begin{array}{c|cc}
\quad & \text{C} & \text{D} \\
\hline
\text{C} & 1 & S \\
\text{D} &  T & 0 
\end{array} \eqcomma
\label{eqn_payoff_matrix}
\end{equation}
meaning that if player $1$ chooses to cooperate and player $2$ cooperates as well, both collect a payoff of $1$. However, if player $2$ defects, player $1$ only collects the payoff 
$S$ and player $2$ collects the payoff $T$, and vice versa. Finally, if both players defect, both obtain no payoff. 

In reality, individuals make many successive strategic decisions and adapt their strategies over time. This behavior is commonly modeled by a replicator dynamics~\cite{evogamesgraphs,Cressman22072014,Helbing1996rep}, in which individuals copy the strategy of a randomly selected neighbor with a probability that depends on the difference of the payoff of the two involved players. In general, individuals tend to copy the strategy of players who have earned a higher payoff compared to themselves. 
Here, we use synchronized updates, meaning that after each round of the game, in which each node plays one game with each of her neighbors and payoffs are distributed according to the aforementioned payoff matrix, every node $i$ chooses a neighbor $j$ at random and copies her strategy with the probability~\cite{PhysRevE.58.69}
\begin{equation}
P_{i \leftarrow j} = \frac{1}{2}\left(1- \tanh\left[ \pi_{i} -\pi_{j} \right] \right) \eqcomma
\label{eqn_pcopy}
\end{equation}
where where $\pi_i$ and $\pi_j$ denote the payoffs of node $i$ and $j$. 

Depending on the values of $S \in [-1,1]$ and $T \in [0,2]$, there are different stable choices of strategies. In the Stag Hunt game, for which we have $S<0$ and $T<1$, we have bistability: both full cooperation as well as full defection are stable stationary solutions. In the Prisoner's Dilemma, i.e. for $S<0$ and $T>1$, full defection is the only stable strategy. In the Snowdrift game, $S>0$ and $T>1$, the only stable solution is an intermediate density of cooperators. Finally, in the Harmony game, $S>0$ and $T<1$, only full cooperation is a stable solution. 

However, in reality, individuals do not make decisions exclusively based on the payoffs of their neighbors. 
Instead, individuals are simultaneously exposed to social influence and hence the opinion of the peer group of an individual cannot be neglected in understanding what drives cooperation in strategic games. Opinions of individuals propagate through a contact network. This behavior is widely described by the voter modelc~\cite{opac-b1095541}, where individuals adopt the opinion of a randomly selected neighbor. We assume that propagating an opinion that is considered anti-social, like defection in our model, is less likely than propagating opinions which are socially accepted, like cooperation in our model. This could be the result of media campaigns or similar measures. We take this effect into account by introducing a bias, $\beta \in [0,1]$, into the voter model. Individuals then adopt the opinion of a randomly selected neighbor with probability $\beta$ if this opinion is cooperation and $1-\beta$ if it is defection. Values of $\beta > 0.5$ hence reflect a positive bias towards cooperation in the opinion dynamics. 
The opinion of an individual can be understood as her proclamation of intend regarding her choice of strategy in the game layer.

As mentioned before, social influence has an impact on the decision of individuals. To mimic this fact, we couple the opinion propagation and game dynamics. In particular, we define a parameter $\gamma$, 
which represents the tendency of individuals to act in agreement with their proclamations, and hence constitutes the coupling strength between social influence and game dynamics.
In particular, at each update step, with probability $\gamma$ a node copies her state from one layer to the other, and with probability $1-\gamma$ updates the strategy in the game layer according to the game dynamics and in the opinion layer according to the biased voter model
(see Fig.~\ref{mod}).

\subsection*{Mixed populations}

In reality, individuals or institutions interact in strategic games via a contact network, like a network between firms or countries. We will discuss the influence of the structure of the contact network and
of the correlations between different networks later. For now, we study the model on a mixed population, in other words, we assume a homogeneous and infinite population in the absence of dynamical correlations and noise. 

The mixed population (meanfield) assumption allows us to derive differential equations for the evolution of the density of cooperators $c_I$ in the game layer and $c_{II}$ in the opinion layer (see Supplementary Materials),
\begin{eqnarray}
\begin{split}
\partial_{t}c_{I} =&(1-\gamma)c_{I}(1-c_{I})\tanh\left[ \left<k\right> \left( c_{I}(1-T)+S(1-c_{I}) \right) \right] \\ & + \gamma (c_{II}-c_{I}) \eqcomma \\
\partial_{t}c_{II} =&(1-\gamma)(2\beta-1)c_{II}(1-c_{II}) + \gamma (c_{I}-c_{II}) \eqcomma
\end{split}
\label{mf_eq}
\end{eqnarray}
where $S$ and $T$ denote the parameters from the payoff matrix, equation~(\ref{eqn_payoff_matrix}), $\gamma \in [0,1]$ controls the strength of the coupling between the opinion and game dynamics and $\beta \in [0,1]$ is the bias of the opinion dynamics.
Finally, $\left<k\right>$ denotes the mean degree of the contact network. 

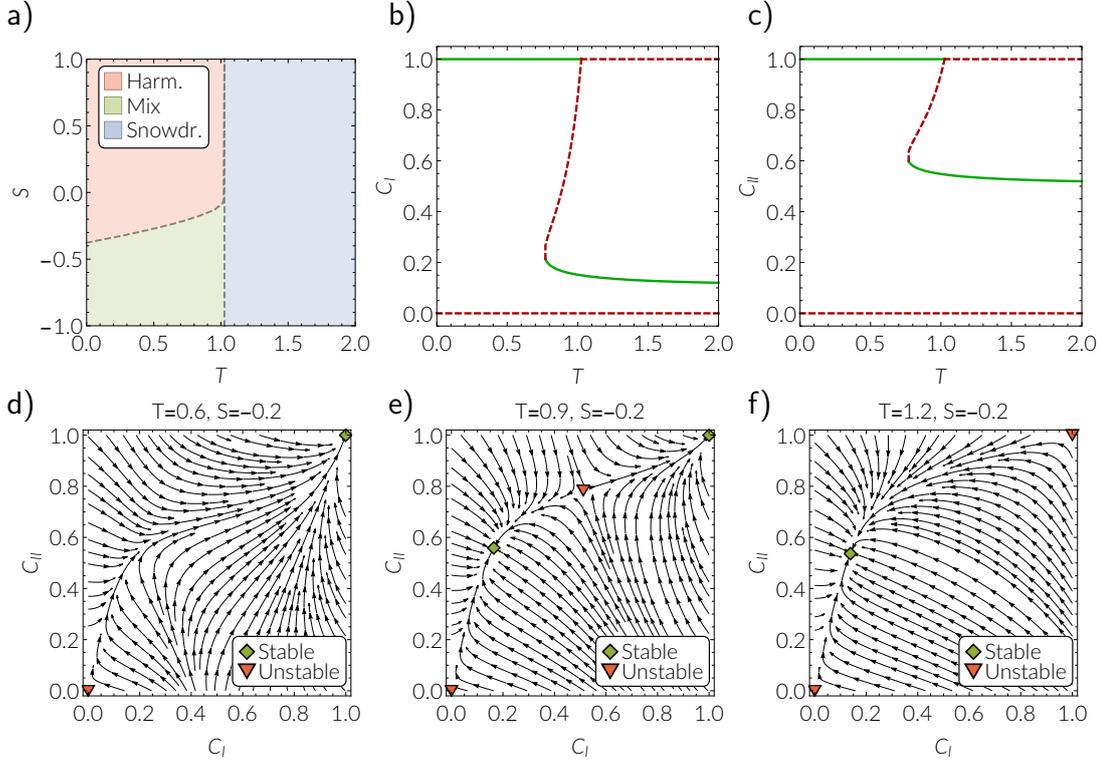
\begin{figure*}
\centering
 \begin{overpic}[width=0.8\textwidth]{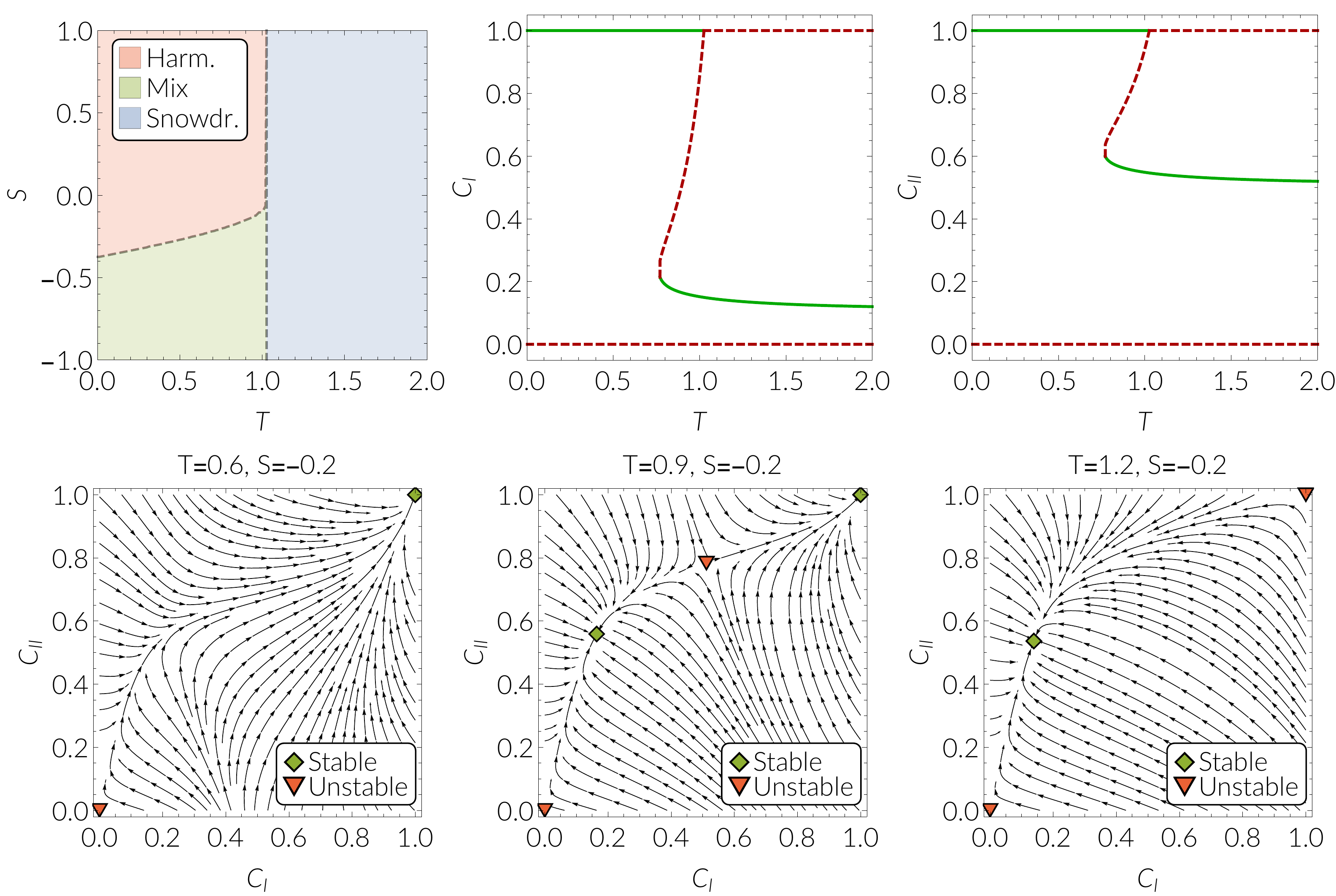} 
 \put (0,68) {\large \textsf{a)}}
  \put (35,68) {\large \textsf{b)}}
    \put (68,68) {\large \textsf{c)}}
     \put (0,32) {\large \textsf{d)}}
  \put (35,32) {\large \textsf{e)}}
    \put (68,32) {\large \textsf{f)}}
\end{overpic}
\caption{
Behavior of the system for $\gamma = 0.2$ and $\beta = 0.7$ and $\left<k\right> = 6$. \textbf{(a)} Shows the phase diagram. In the red area, full cooperation is the only stable solution and we effectively observe the behavior known from the Harmony game. In the opinion layer, ``cooperate'' is the only prevailing opinion. This behavior is illustrated in the stream plot shown in \textbf{(d)} for $T=0.6$ and $S = -0.2$. 
In the blue area in \textbf{(a)}, the only stable solution is a mixed state with $0 < C_I < 1$ and $0 < C_{II} < 1$, and in general we have $C_I \neq C_{II}$. 
This region corresponds to $T > T_{c,2} \approx 1.03$. \textbf{(f)} shows this behavior as a stream plot for $T=1.2$ and $S=-0.2$. 
In the green area in \textbf{(a)}, we have a bistable behavior, where either full cooperation is approached in both layers, but the mixed state is stable as well. \textbf{(e)} shows this behavior as a stream plot for $T=0.9$ and $S=-0.2$. In the bifurcation diagrams in \textbf{(b)} and \textbf{(c)} this region corresponds to $T_{c,1} < T < T_{c,2}$, where $T_{c,1} \approx 0.77$. Green solid lines represent stable fixed points and dashed red lines unstable fixed points. 
\label{fig_mf_grid}
}
\end{figure*}

In the following, we discuss the dynamical properties of the system described by equation~\eqref{mf_eq} for fixed values of $\beta$ and $\gamma$.
We find three regions of different qualitative behavior, depending on the values of parameters $T$ and $S$. 
In particular, we find a region in which the system effectively behaves like the harmony game (red region in Fig.~\ref{fig_mf_grid}a), which means that only full cooperation in both layers is a stable solution (see Fig.~\ref{fig_mf_grid}d). 
Furthermore, we find a region where the system effectively behaves like the snowdrift game (blue region in Fig.~\ref{fig_mf_grid}a). In this region, the only stable solution is a mixed state, where a finite fraction of the population cooperates (see Fig.~\ref{fig_mf_grid}f). In this region, in general, the density of cooperators in the game dynamics and those who proclaim cooperation are not the same. 
Finally, there is a region which can be described as a mixture of the two above cases (green region in Fig.~\ref{fig_mf_grid}a). In this region, the system exhibits a bistable behavior. Full cooperation in both layers is a stable solution as well as a mixed state as described above (see Fig.~\ref{fig_mf_grid}e). 
The bistable region emerges 
as the system undergoes a saddle-node bifurcation. Let us fix $S=-0.2$ and increase the value of $T$. At $T=T_{c,1} \approx 0.77$ the system undergoes a saddle-node bifurcation as a pair of fixed points, one stable and one unstable, appear (see Fig.~\ref{fig_mf_grid}b and c). Increasing $T$ further, at $T = T_{c,2} \approx 1.03$ the system undergoes a
transcritical bifurcation and the solution which corresponds to full cooperation becomes unstable. In the supercritical regime, only the mixed state is stable.    

To sum up, we have shown that the coupling to the opinion dynamics shifts the effective behavior of the game dynamics compared to the isolated case. The coupled system exhibits effectively a Harmony-like behavior, a Snowdrift-like behavior, or a mixture of both. Interestingly, the coupling to the biased opinion dynamics successfully avoids the situation of complete defection. So far, we have considered a fully mixed, homogeneous population. In the following, we discuss the impact of the topology of the underlying contact networks as well as the relationship between the two layers of the system.

\begin{figure*}[t]
\centering
\piclabt{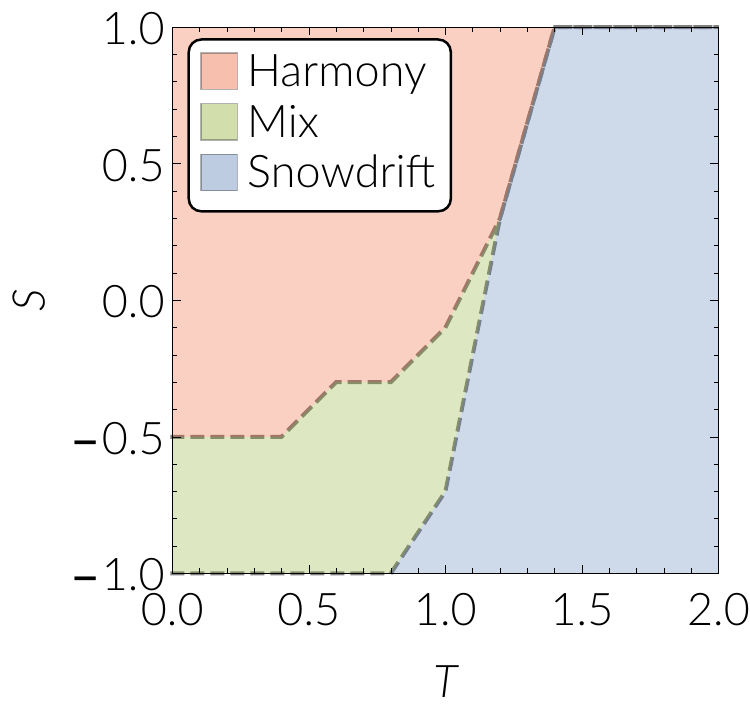}{a}
\piclabt{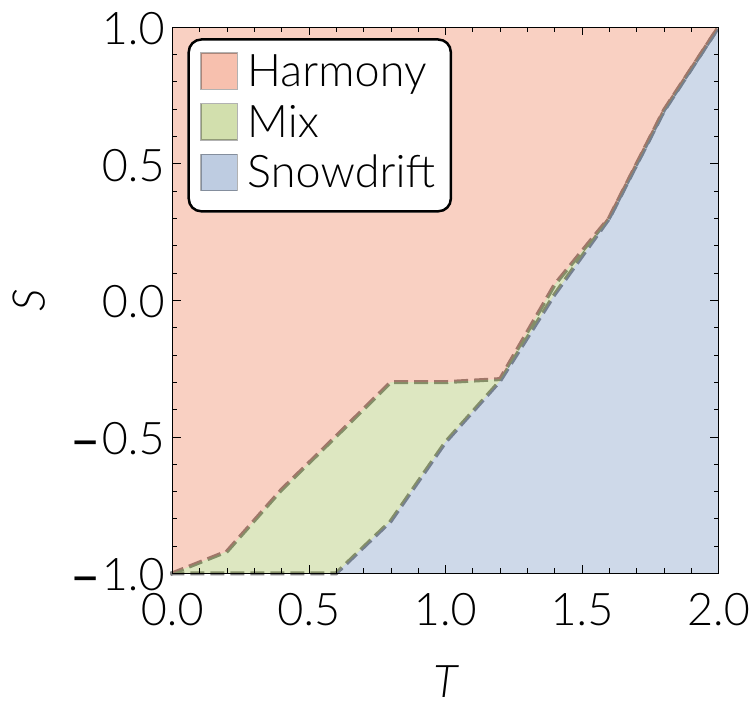}{b} 
\piclabt{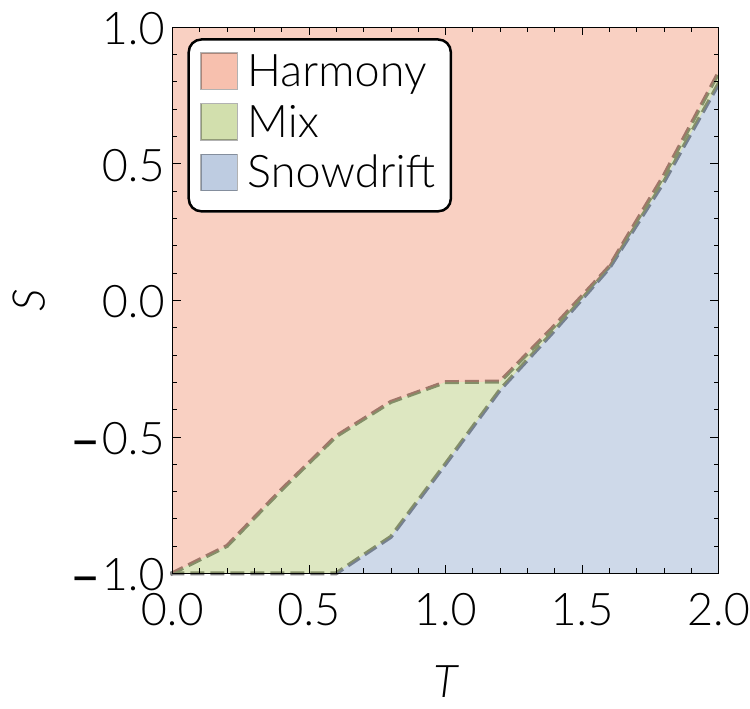}{c}
\begin{overpic}[width=0.22\textwidth]{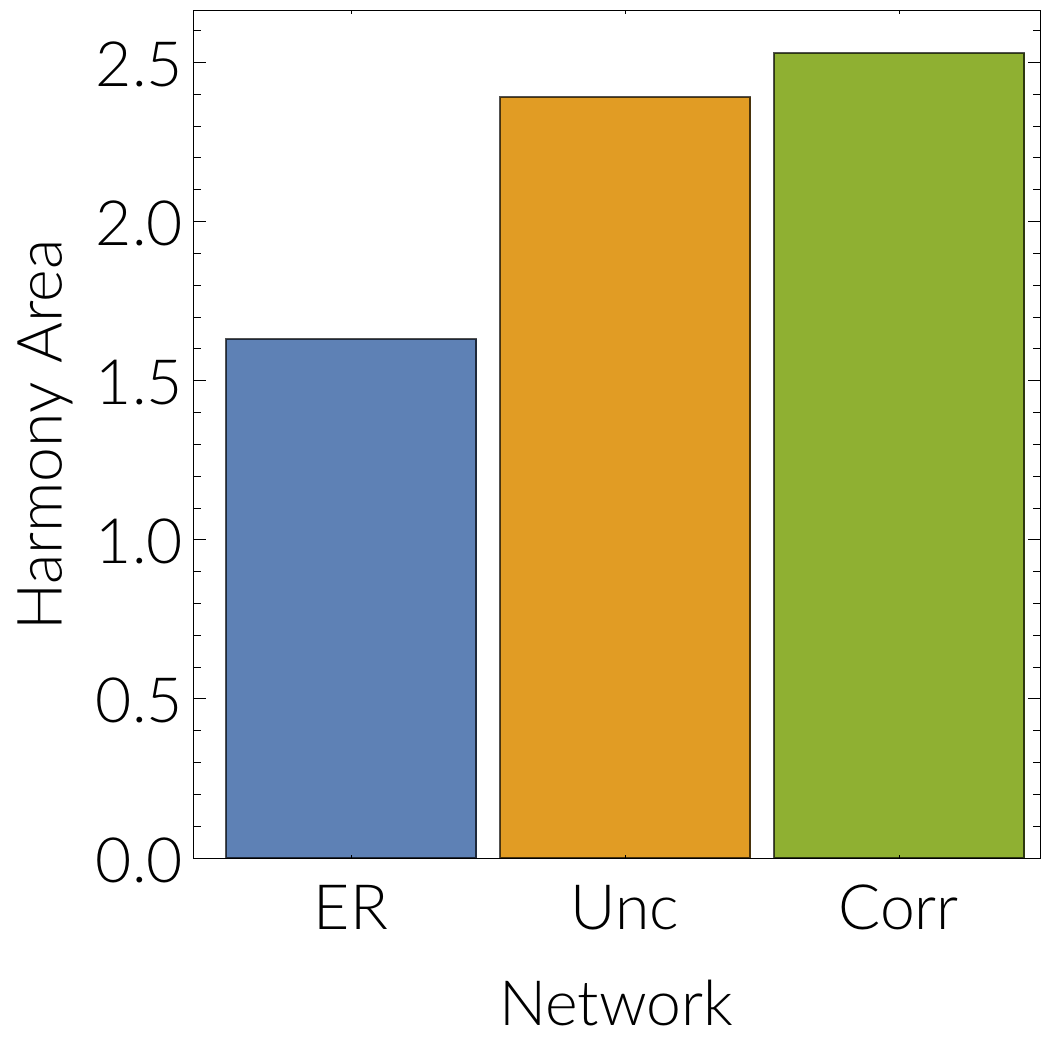}
 \put (3,103) {\large \textsf{d)}}
\end{overpic}
\caption{ 
Approximated phase diagrams (see Supplementary Materials) from numerical simulations.
\textbf{(a)} Erdős Rényi network with mean degree $\left< k \right> = 6$.
\textbf{(b)} Using GMM multiplexes, $\gamma = 0.2$ and $\beta = 0.7$, $N=10000$ nodes, and mean degree $\left< k \right> \approx 6$,  a power-law exponent $2.9$ and temperature $T_\text{GMM}=0.4$. Layers are uncorrelated, i.e. $g=\nu=0$.
\textbf{(c)} same as (b) but with geometric correlation, in particular $g=\nu=1$.
\textbf{(d)} shows the size of the ``Harmony'' area in the phasespace for different parameters, i.e. the size of the red area in (a-c). 
The blue bar is for the Erdős Rényi networks as shown in (a), the yellow bar represent the GMM model without correlations as shown in (b), and the green bar denotes the GMM model with correlations as presented in (c).
\label{fig_phase_num}
}
\end{figure*}
\begin{figure*}[t]
\centering
\includegraphics[width=0.7\linewidth]{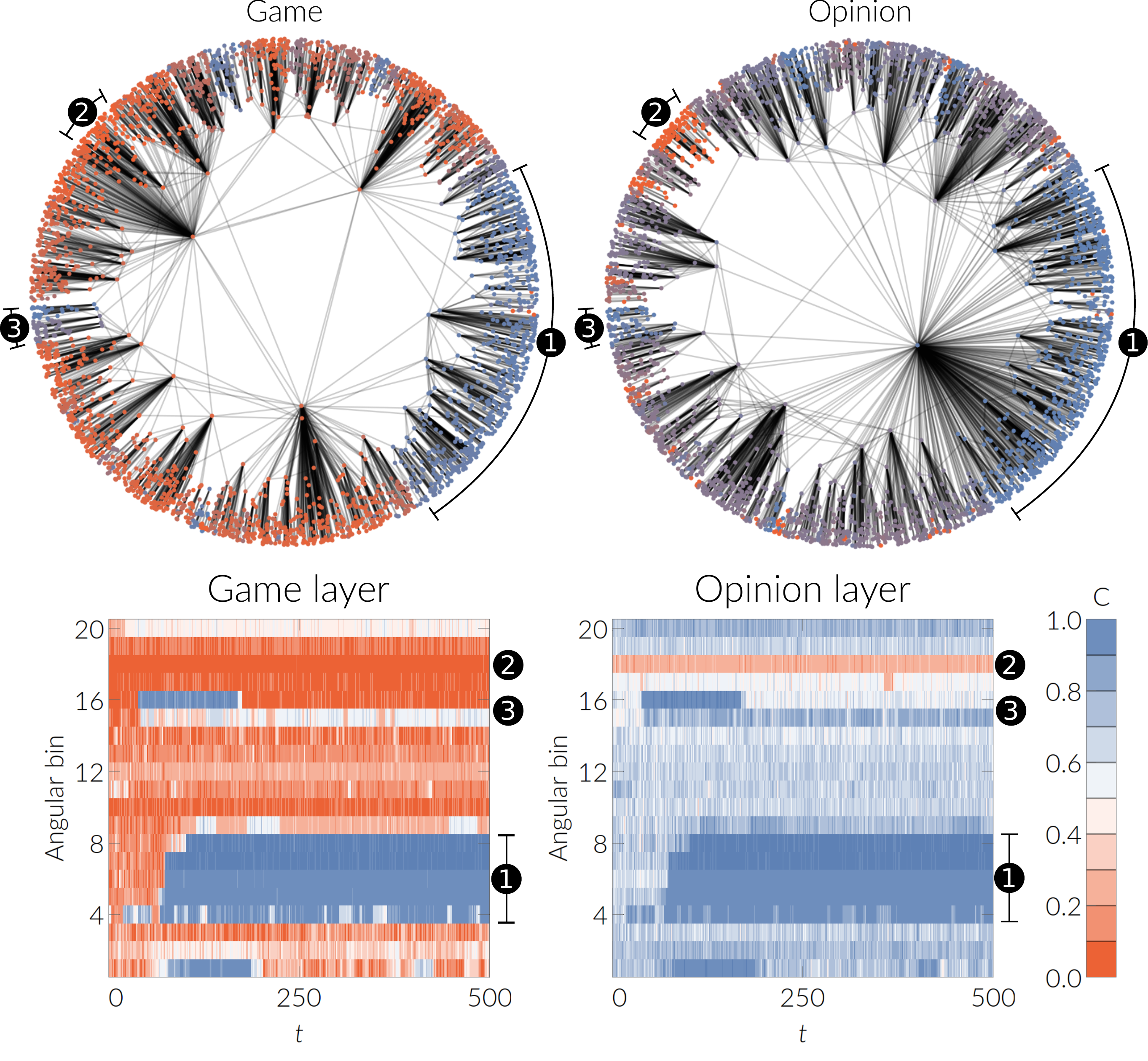}
\caption{
Polarization of the system in the presence of angular correlations between the layers ($g=1,\nu=0$) for a multiplex with $N=5000$ nodes, a power-law exponent $2.9$, temperature $0.2$, and mean degree $6$ in both layers.
Parameters of the game are $T=0.8$ and $S=-0.4$, the bias $\beta$ is $0.7$ and the coupling strength is $0.2$. Results are for a single realization of our model starting with a density of cooperators of $0.1$ in each layer. A visualization of this realization is shown in the Supplementary Video.
The top row shows visualizations of the network layers. Color coded is the mean state of the each node, averaged over time. Each time step denotes $1000$ update steps of each node.
The bottom row shows the evolution of the density of cooperators in each angular bin. Numbers indicate selected clusters of nodes that tend to adopt the same strategy. Each timestep $t$ denotes $10^3$ rounds. 
\label{fig_corr_os}
}
\end{figure*}
\begin{figure}[t]
\centering
\begin{overpic}[width=1\linewidth]{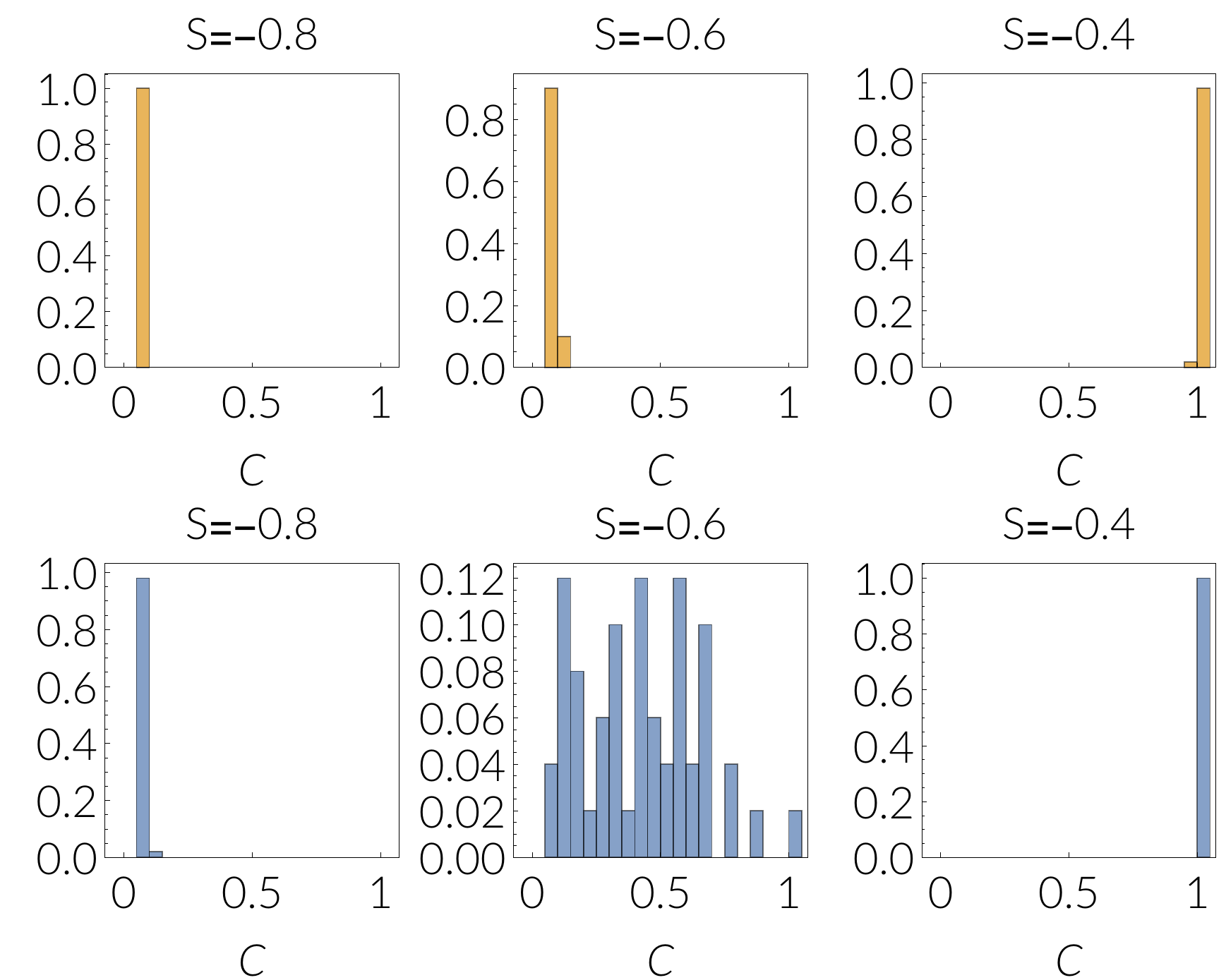}
 \put (4,76) {\large \textsf{a)}}
  \put (36,76) {\large \textsf{b)}}
   \put (70,76) {\large \textsf{c)}}
   \put (4,37) {\large \textsf{d)}}
  \put (36,37) {\large \textsf{e)}}
   \put (70,37) {\large \textsf{f)}}
\end{overpic}
 \caption{
 Distribution of final cooperation (after $5 \times 10^5$ rounds) in the game layer among $50$ realizations of our model. 
 The parameters are $\gamma = 0.2$ and $\beta = 0.7$, $N=10000$ nodes, and mean degree $\left< k \right> \approx 6$. 
 Network layers have a power-law exponent of $2.9$ and temperature $T_\text{GMM}=0.4$. 
 Here, we have fixed $T=0.6$. Plots \textbf{(a)-(c)} show the uncorrelated case ($g=\nu=0$) and \textbf{(d)-(f)} show the case of angular correlations ($g=1, \nu=0$).
 The value of $S$ is shown in the respective plot title. 
 \label{fig_histo}}
\end{figure}

\subsection*{Impact of the structural organization of the multiplex}

Using the assumption of a fully mixed, homogeneous population we have shown how the coupling to the biased opinion dynamics can effectively transform the behavior of the system. However, in reality,
networks are heterogeneous and highly clustered, which can have a significant effect on the outcome of dynamics taking place on the network~\cite{PhysRevLett.95.098104}.

Furthermore, in reality, the social influence layer and the strategic game layer are neither independent nor identical. 
In other words, real multiplex networks are not random combinations of their constituent layer's topologies~\cite{geometry:multilayer}. 
Hence, the contexts---or domains---in which individuals make strategic decisions and by whom they are influenced are related.
In~\cite{geometry:multilayer} the authors have shown that these relations 
are given by geometric correlations in hidden metric spaces underlying each layer of the system. 
These correlations come in two flavors:
popularity correlations, which are correlations between the degrees of nodes, and similarity correlations, which determine how likely an individual is to connect to the same nodes in different layers. In simple terms, these correlations control how ``similar'' the different contexts represented by the layers of the system are. For further details on geometric correlations between layers of real multiplex networks we refer the reader to~\cite{geometry:multilayer}.
Here, we focus on the impact of these structural properties on the dynamics of our model. 
What is, in general, the impact of geometric correlations on the behavior of the system?
In particular, do stronger correlations favor or hinder cooperation? 
To answer these questions, we perform numerical simulation using the geometric multiplex model (GMM) developed in~\cite{geometry:multilayer} (see Methods~\ref{sec_app_gmm}). The model generates networks with a power-law degree distribution and a tunable level of mean local clustering.
Furthermore, we can control the popularity correlations (by tuning parameter $\nu \in [0,1]$) as well as the similarity correlations (by tuning parameter $g \in [0,1]$) independently from the individual layer topologies, which allows us to study their impact in isolation. 
We calculate approximated phase diagrams 
similar to Fig.~\ref{fig_mf_grid}a using the generated networks by performing numerical simulations. In particular, to capture the bistable region of the system, we perform simulations starting from different initial conditions, in particular $C_{I,II} = 0.01$ and $C_{I,II} = 0.99$ respectively. For $\beta=0.7$ and $\gamma=0.2$, the system either reaches full cooperation and consensus (``harmony state''), i.e. $C_I^\text{final},C_{II}^\text{final} = 1$, or a state where a mixed strategy prevails and full consensus is not reached (``snowdrift state''). We find that there is a sharp transition between these regimes (see Fig.1 in the Supplementary Materials). 
We define a critical line, above which the harmony state is reached with a probability of more than $50\%$  (dashed black lines in Fig.2 in the Supplementary Materials). The difference between the critical lines for the different initial conditions is an approximation of the bistable region, in which both the snowdrift state and the harmony state are possible solutions. 
We show the result of this procedure for different structural organizations of the multiplex in Fig.~\ref{fig_phase_num}. 
We observe that heterogeneous and clustered topologies in single layers increase cooperation (compare Figs.~\ref{fig_phase_num}a and b). 
The presence of correlations between the layers increases the region in the parameter space where the harmony solution is approached, and hence further increases cooperation (compare Figs.~\ref{fig_phase_num}b and c). To facilitate this comparison, in Fig.~\ref{fig_phase_num}d we show the size of the ``Harmony'' region in the $T-S$ phasespace.

The impact of correlations in the bistable region is especially interesting.
We find that in this region angular correlations lead to a metastable state in which nodes that adopt the same strategy self-organize into local clusters. These clusters are sets of nodes that are located at small angular distances in the underlying metric space. 
They emerge spontaneously and are metastable in the sense that they can exist for very long times despite the noise present in the system. 
This behavior is shown in Fig.~\ref{fig_corr_os} and in the Supplementary Video. 
The emergence of these clusters can be interpreted as a polarization of society into defecting and cooperating groups. 
Finally, the amount and size of the clusters is highly random and, as a consequence, we observe a broad range of final cooperation densities in this parameter region (see Fig.~\ref{fig_histo}). 

\section*{Discussion}  

Cooperation is common in reality in social dilemmas where many theories predict the prevalence of defection. This contradiction could be resolved by taking into account further domains of interactions between individuals, in particular social influence. 

We have presented a model based on multiplex networks with two layers. One layer represents the domain in which individuals engage in repetitive strategical games. The second layer corresponds to the domain of social influence, which we model using a biased opinion dynamics. In particular, we consider a bias towards cooperative attitudes. We have shown that the coupling between these dynamics can lead to cooperation in scenarios where the pure game dynamics predicts the prevalence of defection. 
Furthermore, we have shown that the coupling of these dynamics in combination with geometric correlations between the layers of the system can lead to a metastable state of high polarization, in which nodes that adopt the same strategy self-organize into local clusters. These findings could explain the emergence and prevalence of polarization observed in many social dilemmas.

Real social and strategic interaction networks evolve in time, and their evolution could depend on the strategic choices of individuals~\cite{our:model,Gulys2015}. Hence, the inclusion of an evolving and adaptive topology constitutes and interesting task for future work. Furthermore, one could include the competition~\cite{ecology20,worldmodel,gom15} between different strategic networks, or incorporate external noise~\cite{Helbing2009pnas}.
Finally, our findings suggest that hidden geometric correlations between different layers of multiplex networks can alter the behavior of the 
dynamics taking place of top of them
significantly, and hence such correlations should be taken into account in future research on dynamical processes on multiplex networks~\cite{geo:targeted}.

\appendix
\section*{Methods}

\subsection*{Geometric multiplex model (GMM)}
\label{sec_app_h2}
\label{sec_app_gmm}

The geometric multiplex model is based on the (single-layer) network construction procedure of the newtonian $\mathbb{S}^{1}$~\cite{Serrano2008} and hyperbolic $\mathbb{H}^{2}$~\cite{Krioukov2010} models. The two models are isomorphic and here we present the results for the $\mathbb{H}^{2}$ version. 
The construction of a network of size $N$ proceed firsts by assigning to each node $i=1,\ldots, N$ its popularity and similarity coordinates $r_i, \theta_i$ and subsequently, connecting each pair of nodes $i, j$ with probability $p(x_{ij})=1/(1+e^{\frac{1}{2T}(x_{ij}-R)})$, where $x_{ij}$ is the hyperbolic distance between the nodes and $R \sim \ln{N}$ (see Supplementary Materials). The connection probability $p(x_{ij})$ is the Fermi-Dirac distribution where the 
\emph{temperature} parameter $T_\text{GMM}$ controls the level of clustering in the network~\cite{Dorogovtsev10-book}. The average clustering $\bar{c}$ is maximized at $T=0$, linearly decreases to zero with $T \in [0,1)$, and is asymptotically zero if $T>1$. As $T \to 0$ the connection probability becomes the step function $p(x_{ij}) \to 1$ if $x_{ij} \leq R$, and $p(x_{ij}) \to 0$ if $x_{ij} > R$. It has been shown that the $\mathbb{S}^{1}$ and $\mathbb{H}^{2}$ models can build synthetic networks reproducing a wide range of structural characteristics of real networks, including power law degree distributions and strong clustering~\cite{Serrano2008,Krioukov2010}.  
The use of these models for the single-layer networks allows for radial and angular coordinate correlations across the different layers. The level of these correlations can be controlled by model parameters $\nu \in [0,1]$ and $g \in [0,1]$, without affecting the topological structure of the single layers. 
The radial correlations, related to the node's degree, increase with parameter $\nu$---at $\nu=0$ there are no radial correlations, while at $\nu=1$ radial correlations are maximized. Similarly, the angular correlations increase with parameter $g$---at $g=0$ there are no angular correlations, while at $g=1$ angular correlations are maximized.
See~\cite{geometry:multilayer} for details.

Finally, we extract the mutually connected component of the system to avoid disconnected nodes. 

\begin{acknowledgments}
K-K. K. acknowledges support by the ERC Grant ``Momentum'' (324247).
R.A. and A.D.-G. acknowledge financial support from MINECO, Projects FIS2012-38266 and FIS2015-71582, and from Generalitat de Catalunya Project 2014SGR-608.
R.A. and A.D.-G. acknowledge support by LASAGNE.
\end{acknowledgments}

\end{document}